\documentclass[useAMS,usenatbib]{mn2e}
\usepackage{myaasmacros,graphicx}

\def\ltsima{$\; \buildrel < \over \sim \;$}
\def\simlt{\lower.5ex\hbox{\ltsima}}   
\def\gtsima{$\; \buildrel > \over \sim \;$}
\def\simgt{\lower.5ex\hbox{\gtsima}}
\newcommand\bcite[1]{\citeauthor{#1} \citeyear{#1}}

\title[Dynamical friction in constant density cores]
{Dynamical friction in constant density cores: a failure of the Chandrasekhar formula}

\author[Read et al.]{J. I. Read$^1$\thanks{Email:
    justin@physik.unizh.ch}, Tobias Goerdt$^1$, Ben Moore$^1$, A. P. Pontzen$^2$, Joachim Stadel$^1$ \& \and George Lake$^1$\\
\\$^1$Institute of Theoretical Physics, University of Z\"urich,  
Winterthurerstrasse 190, 8057 Z\"urich, Switzerland
\\$^2$Institute of Astronomy, Cambridge University, Madingley Road, 
Cambridge, CB3 OHA, England}

\begin{document}

\maketitle

\begin{abstract}
Using analytic calculations and N-body simulations we show that in
constant density (harmonic) cores, sinking satellites undergo an initial
phase of very rapid (super-Chandrasekhar) dynamical friction, after
which they experience no dynamical friction at all. For density profiles with a central power law profile, $\rho \propto r^{-\alpha}$, the infalling satellite heats the background and causes $\alpha$ to decrease. For $\alpha < 0.5$ initially, the satellite generates a small central constant density core and stalls as in the $\alpha = 0$ case. 

We discuss some astrophysical applications of our results to decaying satellite orbits, galactic bars and mergers of supermassive black hole binaries. In a companion paper we show that a central constant density core can provide a natural solution to the timing problem for Fornax's globular clusters.
\end{abstract}


\section{Introduction}\label{sec:introduction}
In a seminal paper, \citet{1943ApJ....97..255C} showed that a massive
particle moving through an infinite, homogeneous and isotropic background of lighter particles  experiences a force of dynamical friction given by:

\begin{equation}
M_c \frac{dv}{dt} = -4\pi G^2 M_c^2
\frac{v}{|v|^3}\ln\left(\frac{b_\mathrm{max}}{b_\mathrm{min}}\right)\int_0^{|v|}M(v')dv'
\label{eqn:chandrasekhar}
\end{equation}
where $M_c$ and $v$ are the mass and velocity of the in-falling
particle, $M(v')dv'$ is the mass density of background objects with
speeds $v' \rightarrow v'+dv'$, and $b_\mathrm{max}$ and $b_\mathrm{min}$ are the
maximum and minimum impact parameters for the encounters\footnote{Note
  that $b_\mathrm{min} \rightarrow 0$ can be achieved (see e.g. \bcite{1976MNRAS.174..467W}; \bcite{1987gady.book.....B} p. 423), while
  $b_\mathrm{max} \rightarrow \infty$ cannot. This is because the derivation
  of equation \ref{eqn:chandrasekhar} assumes an {\it infinite}
  background; $b_\mathrm{max}$ defines a scale on which the infinite
  background should be truncated. This is often, reasonably, take to
  be the radius at which the mean density falls by a factor of two or
  so.}. From here on, we refer to the massive in-falling object as a `Globular Cluster' (GC) and the background of lighter particles as simply `particles'. However, we could equally refer to, for example, a bar moving in a background of stars and dark matter. 

While equation \ref{eqn:chandrasekhar} is only strictly valid for
an infinite, homogeneous and isotropic background, it has been shown
to work remarkably well for satellites orbiting in spherical galaxies
with more general background distributions\footnote{When
corrected for velocity anisotropies, it has also been shown to work
well in aspherical systems (see e.g. \bcite{1977MNRAS.181..735B},
\bcite{1991ApJ...375..544S} and \bcite{2004MNRAS.349..747P}).} (see
e.g. \bcite{1983ApJ...274...53W}, \bcite{1988MNRAS.235..289Z},
\bcite{1997MNRAS.289..253C} and \bcite{1987MNRAS.224..349B}). Such successes make equation \ref{eqn:chandrasekhar} of great practical value. But they beg the question: why has it been so successful, even when it is used so far beyond its expected regime of validity? Are we missing important physical insight into the dynamical friction process in spherical systems? Does Chandrasekhar fail to work well in some situations?

In order to address some of these issues, \citet{1984MNRAS.209..729T} (hereafter TW84) and \citet{1986ApJ...300...93W} (hereafter W86) formulated a perturbative theory of dynamical friction which could be applied to spherical systems. Notice from equation \ref{eqn:chandrasekhar}, that most of the dynamical friction originates from particles with large impact parameters: it is the accumulation of many long range {\it small} interactions which leads to most of the dynamical friction; not the large angle scattering of close encounters. This is why perturbative methods can be used. TW84 and W86 consider a general, small, perturbation to a single background particle; and then sum over all particles in the system to obtain to total torque induced on the perturber. 

In section \ref{sec:analytic}, we will briefly summarise the essence of this perturbation method. For now, it is important to note the key assumptions in the method; and the key results. The two main assumptions are: (i) that the perturbation is small and (ii) that the frequency of the perturber changes with time, $\Omega_s = \Omega_s(t)$, {\it faster than the perturbation can grow non-linear}. For most potentials of interest this second assumption is satistifed. The perturber (the GC) will lose angular momentum to the background particles as a result of the dynamical friction, and $\Omega_s$ will then increase as the GC falls inwards. 

Under the above assumptions, the perturbation method gives us new physical insight into the dynamical friction problem. To the order of the perturbation approximation, {\it all} of the torque comes from background particles which are close to resonance with the perturber. Non-resonant particles do not contribute to the friction at all. This is a key difference between the perturbation solution and that of equation \ref{eqn:chandrasekhar}. It suggests that if equation \ref{eqn:chandrasekhar} is ever going to fail, it would do so for background particle distributions which are especially resonant. 

In this paper, we describe such a super-resonant potential: that of the constant density (harmonic) core. For this special potential, all particles and the perturber always move with constant angular frequency, $\Omega$. In this case, perturbation methods can no longer be used. This is because $\Omega_s = \mathrm{const}$; assumption (ii), above, is violated; and the perturbations, which are always driven at the same resonant frequency, can grow indefinitely\footnote{This is true for {\it any} perturbative scheme (e.g. \bcite{1999ApJ...525..720C}).}.

To cope with this special case, we develop a non-perturbative analytic model using a 3D driven harmonic oscillator. This essentially generalises an earlier result derived by \citet{1972gnbp.coll.....L}. Using our analytic model and N-body simulations, we show that in constant density cores, equation \ref{eqn:chandrasekhar} fails. Sinking satellites undergo an initial phase of very rapid (super-Chandrasekhar) dynamical friction, after which they experience little or no dynamical friction at all. 

\citet{2005astro.ph..8166W} and \citet{2006astro.ph..1138W}, find similar stalling results for galactic bars (which may be thought of as two diametrically opposed satellites) inside constant density cores.

Constant density cores have recently become interesting in astrophysics. Observations of galaxies on all scales from dwarf spheroidals in the Local Group, up to giant spirals suggest that their central dark matter density has such a constant density core on the scale of $\sim 1$\,kpc (see e.g. \bcite{2003ApJ...588L..21K}, \bcite{2001ApJ...552L..23D}, \bcite{2001MNRAS.323..285B} and \bcite{2001MNRAS.327L..27B}); but see also \citet{2004MNRAS.355..794H} and \citet{2004ApJ...617.1059R},  for a discussion of the potential systematic errors in such observations. If cores are present at the centre of galaxies, their resonant properties can significantly affect the dynamics. Bars can be much longer lived\footnote{\citet{1998ApJ...493L...5D} and \citet{2000ApJ...543..704D} show that low central dark matter densities lead to bars which remain fast. Here we discuss the extreme case of constant density cores, in which we show that bars would not slow down at all. This agrees well with earlier findings by \citet{2005astro.ph..8166W} and \citet{2006astro.ph..1138W}.}, while in-falling satellites and GCs will stall at the core radius. In a companion paper, \citet{Goerdt:2006rw}, we investigate this last idea further (see also \bcite{1998MNRAS.297..517H} and \bcite{2006astro.ph..1490S}). The Fornax dwarf spheroidal galaxy in the Local Group has 5 GCs at a range of projected radii. Application of Chandrasekhar dynamical friction suggests the clusters should rapidly fall to the centre of Fornax from their current positions; fine tuning is required to have them arrive at their present positions at the current epoch. In \citet{Goerdt:2006rw} we show that a small core of radius greater than 0.24\,kpc can solve this problem by causing some, or all, of Fornax's GC to stall.

This paper is organised as follows: in section \ref{sec:analytic} we briefly review the perturbative method for calculating dynamical friction and demonstrate that it fails for the special case of a constant density core. We show that insight can be gained from a non-perturbative approach by modelling the system as a driven harmonic oscillator. In section \ref{sec:simulations} we describe our semi-analytic and full N-body simulations. In section \ref{sec:results}, we test our analytic model against these high resolution ($\sim 10^7$ particles) simulations of satellites sinking in harmonic cores. We demonstrate that such high resolution is required in order to reduce numerical precession of the GC orbit plane, but that near-converged results for the GC orbit can be obtained at lower resolution with O($10^6$ particles). We discuss the importance of the initial GC orbit, mass, the underlying gravitational potential and the particle-particle interactions. Finally, in section \ref{sec:conclusions} we briefly discuss the implications of these results and present our conclusions.

\section{Analytic results}\label{sec:analytic}

\subsection{A brief review of the perturbation method}\label{sec:pertrev}

The essence of the TW84 perturbative approach to dynamical friction can be understood in the following way: consider a spherical potential, $\Phi(r)$, to which a small {\it non-axisymmetric} perturbation, $\Phi_s$, is applied. The perturbation rotates with angular frequency $\Omega_s$. In this case, the equations of motion of a test particle moving in a frame stationary with respect to the perturbation are given by:

\begin{equation}
\underline{\ddot{r}} + \underline{\nabla}[\Phi+\Phi_s] + 2\underline{\Omega}_s \times \underline{\dot{r}} + \underline{\Omega}_s \times (\underline{\Omega}_s \times \underline{r}) = 0
\label{eqn:wbfd}
\end{equation}
where the third and fourth terms are the familiar Coriolis and Centrifugal inertial forces respectively. 

The problem is symmetric about the plane containing the perturbation, so it makes sense to work in cylindrical coordinates: $\underline{r} = \underline{r}(R,\phi,z)$. Equation \ref{eqn:wbfd} then reduces to: 

\begin{equation}
\ddot{R} - R\dot{\phi}^2 = - \frac{\partial [\Phi+\Phi_s]}{\partial R} + 2R\dot{\phi}\Omega_s+\Omega_s^2 R
\label{eqn:wbfdcylin1}
\end{equation}
\begin{equation}
R\ddot{\phi} + 2\dot{R}\dot{\phi} = - \frac{1}{R}\frac{\partial\Phi_s}{\partial \phi}-2\dot{R}\Omega_s
\label{eqn:wbfdcylin2}
\end{equation}
and equation \ref{eqn:wbfdcylin2} can be rearranged to give: 
\begin{equation}
\frac{d}{dt}\left(R^2\dot{\phi}\right) = \dot{J}_z = -\frac{\partial \Phi_s}{\partial \phi} - 2 R\dot{R}\Omega_s 
\label{eqn:Jzdot}
\end{equation}
where $J_z$ is the $z$-component of the specific angular momentum of the test particle and we have introduced the notation $\Omega_s = |\underline{\Omega}_s|$; and similarly for other vectors.

In order to solve equation \ref{eqn:Jzdot}, we must now specify the perturbation, $\Phi_s(R,\phi)$, and the angular motion of the test particle, $\phi(t)$. While it is not necessary in general, it also greatly simplifies the analysis to assume that $\dot{R}=0$, which we do from here on. With this assumption, we can still illustrate usefully the key points of the perturbation method. 

We consider the perturbation: $\Phi_s = A e^{im\phi}$. This is instructive since it is then one component of a more general Fourier series sum. We can find $\phi(t)$ if we assume that the perturbation is small. The usual trick is to suppose that over short times the particle trajectory is the same as in the unperturbed case. For the {\it unperturbed} case, $\Phi_s, \Omega_s \rightarrow 0$ and equation \ref{eqn:Jzdot} gives: $\phi_{in} = \Omega_*t + \mathrm{const}$; where the subscript $in$ reminds us that this is now with respect to an inertial frame. Transforming $\phi_{in}$ to the non-inertial frame rotating with $\Omega_s$, gives: $\phi = (\Omega_*-\Omega_s)t + \mathrm{const}$.

Equation \ref{eqn:Jzdot} may now be integrated to give:

\begin{equation}
J_z = -\mathrm{Re}\left\{\frac{A\exp[im(\Omega_* - \Omega_s)t]}{(\Omega_* - \Omega_s)}\right\} - R^2 \Omega_s
\label{eqn:Jz}
\end{equation} 

It is clear from equation \ref{eqn:Jz} that $J_z$ just oscillates with no time averaged change\footnotemark (i.e. no dynamical friction) unless $\Omega_* = \Omega_s$. At this {\it resonant} frequency the test particle appears to have a pathological specific angular momentum. In practise this just means that the approximation that the perturbation is small fails. 

\footnotetext{Recall that we have assumed that $R=\mathrm{const.}$}

TW84 show that if $\Omega_s = \Omega_s(t)$, then this problem can be solved. Provided $\Omega_s$ changes faster than the time taken for the perturbation to grow into the non-linear regime, then we can sum over all of the resonant interactions from the background particles and calculate the resulting torque on the perturber\footnote{Note that there is now an extra term which should also be included in equation \ref{eqn:wbfd}: $\underline{\dot{\Omega}}\times \underline{r}$; we assume that this is small.}. There are two regimes of interest: fast and slow passages through resonance. The fast passage through resonance recovers the LBK torque formula \citep{1972MNRAS.157....1L}. This is the perturbation theory equivalent of equation \ref{eqn:chandrasekhar}: it describes the dynamical friction. For slow passages through resonance, TW84 find quite different behaviour. The torque is stronger than in the LBK case, reversible, and can lead to the capture (gravitational binding) of background particles by the perturber. These differences led TW84 to refer to this as {\it dynamical feedback}, rather than friction. We return to this effect in section \ref{sec:nonpert}.

In this paper we discuss a special potential of interest generated by a constant density core. For this special case, the divergence in equation \ref{eqn:Jzdot} persists because $\Omega_s$ stays fixed. The potential for a constant density core is the harmonic potential given by:

\begin{equation}
\Phi = \frac{\Omega^2}{2}r^2 + \mathrm{const}
\label{eqn:harmpot}
\end{equation}
where $\Omega$ is the angular frequency of test particles (including the GC; $\Omega_s=\Omega$) in the harmonic core.

The equation of motion for the GC perturber moving in the harmonic potential is given by:

\begin{equation}
\underline{\ddot{r}}_c + \underline{\nabla}\Phi = 0 = \underline{\ddot{r}}_c + \Omega^2 \underline{r}_c
\label{eqn:sateqmotion}
\end{equation}
Which may be trivially solved to give the general solution:

\begin{equation}
\underline{r}_c = [X\sin(\Omega t+\phi_x),Y\sin(\Omega t + \phi_y)]
\label{eqn:satmotion}
\end{equation}

From equation \ref{eqn:satmotion} we can see that orbits in harmonic potentials are of fixed relative phase angle, closed and of constant angular frequency, $\Omega$. This means that provided the potential remains harmonic, any perturbation to the GC orbit - including dynamical friction and loss of angular momentum - will not change $\Omega$ or $\Omega_s = \Omega$. In other words, $\Omega_s \neq \Omega_s(t)$ and we can no longer apply perturbation theory methods. 

M. Weinberg (private communication) has made the valid point that the perturber itself, and the non-spherically symmetric background distribution it induces, cause deviations from true harmony. It may be possible to use a perturbative approach in this, more realistic, case.

\subsection{A non-perturbative approach}\label{sec:nonpert}

Perturbation methods fail for the harmonic core. However, all is not lost analytically. We can still gain much insight by writing down the equations of motion for the GC and a tracer background population, and searching for stable solutions. As we shall show next, for the special case of a harmonic potential plus point mass perturber (the GC), solutions exist where the background particles rotate about the GC on stable epicycles. Stable orbits mean no time averaged angular momentum transfer and, therefore, no dynamical friction. 

Such a model allows us to make firm qualitative (if not quantitative) statements about what will happen when a GC is introduced to an isotropic constant density core. Initially, particles will be in equilibrium in the constant density core. As the GC approaches the core, the system will need to rearrange itself and reach a new equilibrium state. The non-linear interplay between the GC and the background distribution during this rearrangement leads to a period of enhanced, super-resonant friction. After $\sim 1$  dynamical time, the distribution function of the background will now be the correct one for the GC plus harmonic core, and dynamical friction will cease. Note that this rearrangement may also be understood in terms of the TW84 {\it dynamical feedback} discussed in section \ref{sec:pertrev}.

Our model does not include the back-reaction of the test particles on the GC, nor does it include the interaction between the background particles themselves. However, we find a good agreement between our analytic model and full N-body simulations, which include the above effects, in section \ref{sec:simulations}. This suggests that our simple model does capture the essential physics of the problem.

\begin{figure}
\begin{center}
\includegraphics[width=4cm,angle=-90]{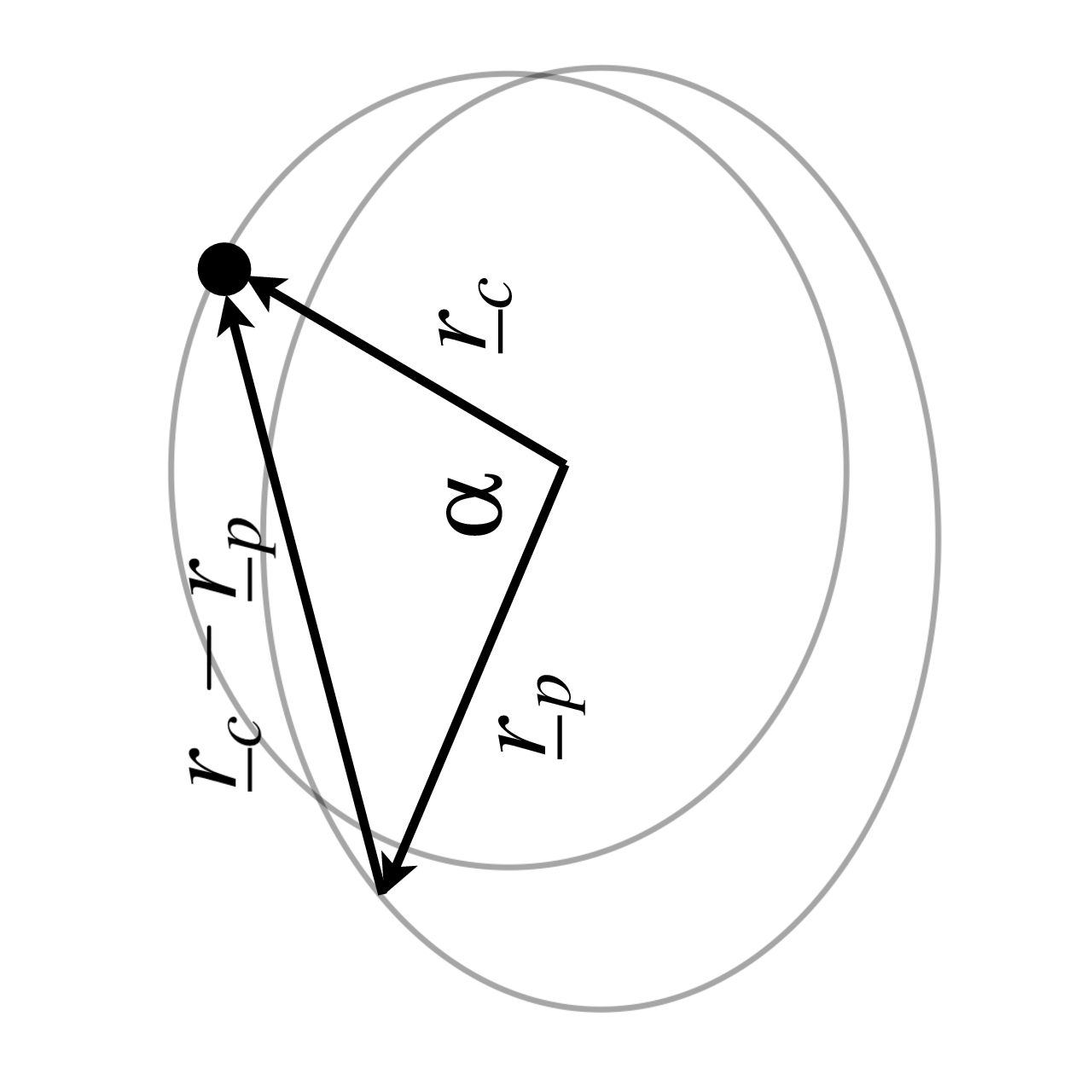}
\caption{A schematic diagram of the analytic set-up. The in-falling GC is marked by the solid black circle. The GC and particle orbits are marked by the grey ellipses. See text for further details.}
\label{fig:analytic}
\end{center}
\end{figure}

The analytic set-up is shown in Figure \ref{fig:analytic}. The in-falling GC at a radius, $\underline{r}_c$ is marked by the black circle and is a phase angle, $\alpha$, away from a given background particle at a radius, $\underline{r}_p$. We assume that the underlying potential is always harmonic (given by equation \ref{eqn:harmpot}); the GC is well approximated by a point mass; and the background potential is nailed down (this is reasonable provided that $M_c \ll M_{en}$ where $M_{en}$ is the mass enclosed by the GC). Under these assumptions, the equation of motion for a single background, massless {\it tracer}, particle is given by:

\begin{equation}
\underline{\ddot{r}}_p + \Omega^2 \underline{r}_p = \underline{F} = \frac{G M_c (\underline{r}_c - \underline{r}_p )}{|\underline{r}_c - \underline{r}_p|^3}
\label{eqn:starmotion}
\end{equation}
where $\underline{F}$ is the specific force on the particle from the GC, $G$ is the gravitational constant and $M_c$ is the mass of the GC. 

We now search for stable solutions to equation \ref{eqn:starmotion} where the GC orbit is unchanged by the background. Combining equations \ref{eqn:sateqmotion} and \ref{eqn:starmotion} gives:

\begin{equation}
\underline{\ddot{r}}_d + \left(\Omega^2 - \frac{GM_c}{|\underline{r}_d|^3}\right) \underline{r}_d = 0
\label{eqn:dsol}
\end{equation}
where $\underline{r}_d = \underline{r}_c -\underline{r}_p$. 

From equation \ref{eqn:dsol}, it is clear that stable solutions exist where the background particles move on circular epicycles about the GC with $|\underline{r}_d| = \mathrm{const}$. However, more general solutions may be found by noting that equation \ref{eqn:dsol} is spherically symmetric. Moving to spherical polar coordinates: $\underline{r}_d = (r,\theta,\phi)$, it is straightforward to show that the $\theta$ and $\phi$ specific angular momenta are conserved: $J_\theta = r^2\dot{\theta} = \mathrm{const};  J_\phi = \mathrm{const}$. This is to be expected given the symmetry of the problem. Equation \ref{eqn:dsol} then reduces to:  

\begin{equation}
\ddot{r} + \frac{d\Phi_\mathrm{eff}}{dr} = 0
\label{eqn:r}
\end{equation}
\begin{equation}
\Phi_\mathrm{eff} = \frac{\Omega^2 r^2}{2} + \frac{GM_c}{r} + \frac{J_\theta^2}{2r^2}
\label{eqn:phieff}
\end{equation}
where $\Phi_\mathrm{eff}$ is the effective potential. 

From equation \ref{eqn:r}, we can see that in general, the background particles move on epicycles about the GC. These epicycles will not be closed, but they are quasi-periodic. Provided the distribution function of these background particles is the correct equilibrium distribution for the GC plus the harmonic core, there will be no time averaged momentum exchange between the GC and the background, and therefore no dynamical friction. The epicyclic orbits are stable, as can be readily seen by considering $\frac{\partial^2 \Phi_\mathrm{eff}}{\partial r^2}$.

The existence of stable analytic solutions is a very special property of the harmonic potential; they exist because $\Omega = \mathrm{const}$. In more general potentials, $\Omega = \Omega(\underline{r}_p)$ and the symmetry of equation \ref{eqn:dsol} is broken. This is an important issue. One can imagine a thought experiment where a GC is held (artificially) on a fixed orbit in a general spherical potential. After a few dynamical times, it will have scattered the resonant background particles, reducing the torque from the background to zero. It is important to stress that this is quite different to the situation we have described in this paper. In the above thought experiment, a tiny perturbation to the GC orbit (which must in practise occur as a result of its self-consistent interaction with the background) will expose the GC to an entirely new set of resonant background particles: dynamical friction will not cease. In our example, however, {\it any} perturbation to the GC orbit will not alter its orbital frequency at all: the resonances will remain unchanged. This is why our assumption, above, that the GC orbit is fixed is not an important one for the harmonic potential, but would be for any other potential. We test that this is indeed the case by relaxing the assumption of a fixed GC orbit in section \ref{sec:simulations}.

We can use the above solution to calculate the final distribution of background particles at equilibrium when the dynamical friction ceases. The key point is that the final distribution will move on stable epicycles about the GC. Firstly, this means that we can expect a density enhancement around the GC, and a depletion of particles away from the GC. Secondly, we can expect a large depletion in counter-rotating particles with respect to the centre of the potential. All particles, whether they move on co- or counter-rotating epicycles have guiding centres which co-rotate with the GC. For $|\underline{r}_d|<|\underline{r}_c|$, the radius of the epicyclic orbit is smaller than that of the GC: none of these particles can counter-rotate with respect to the centre of the potential. For $|\underline{r}_d|>|\underline{r}_c|$, particles on counter-rotating epicycles can appear to counter-rotate with respect to the centre of the potential. These will be a small fraction of the total particles which remain. We test these qualitative expectations, using simulations, in section \ref{sec:simulations}.

A final point, which will become important later on, is that the orbit plane of the GC matters. Equation \ref{eqn:dsol} is spherically symmetric about the GC and hides this fact. If the GC orbit changes (and noise within the full N-body simulations can cause this to happen) then the angular frequency vector of the GC, $\underline{\Omega}$, will change: the background distribution will no longer be in equilibrium with the GC. The system will have to move once again into equilibrium and this rearrangement will lead to some associated dynamical friction on the GC. 

\subsection{The Kalnajs solution}

Our analytic method is a more general case of an earlier result found by \citet{1972gnbp.coll.....L}. Kalnajs studied dynamical friction in a uniformly rotating sheet in which all particles initially move on circular orbits. This is an equivalent problem to a GC moving on a circular orbit within a harmonic potential. He showed, using results from plasma physics, that in this case dynamical friction will vanish. Here we generalise this result to a GC moving on a general orbit within a harmonic potential. In our solution, the background perturbation need not lie in the plane of the GC orbit. 

\section{Simulations}\label{sec:simulations}

In this section we compare semi-analytic and full N-body simulations to the analytic formulae derived in section \ref{sec:analytic}. The simulations are labelled as in Table \ref{tab:simulations} and described in detail in the sub-sections below. 

The analytic arguments given in section \ref{sec:analytic} suggest that once a GC is introduced to a constant density background, the system will move towards a stable equilibrium where the background particles move on epicycles about the GC. However, this simple analytic argument cannot say anything about interactions between the GC and the background particles prior to such an equilibrium being achieved; or of interactions between the background particles themselves. In this section we investigate this approach to equilibrium using numerical simulations. We use two types of simulation. The semi-analytic (SA) run solves equation \ref{eqn:starmotion} numerically. We still assume that the GC orbit is fixed, however we can study how the system moves from one equilibrium state (without the GC) to its final equilibrium with the GC. The full N-body (NB) run includes the interaction between the particles and the GC self-consistently. The GC is now free to respond to the background particles. This allows us to study the full effect of dynamical friction on the GC as the system moves towards its new equilibrium state. We compare results for a GC initially on a circular orbit and an elliptical orbit. 

\begin{table*}
\begin{center}
\setlength{\arrayrulewidth}{0.5mm}
\begin{tabular}{llllll}
\hline
{\it Simulation} & {\it Description} & {\it GC orbit} & {\it Potential} & {\it Resolution} & $M_c$ \\
\hline
SA\{c,e\} & Semi-analytic & Fixed, \{(c)irc., (e)llip.\} & Fixed, harmonic & $10^5$ tracer & 
$2 \times 10^5$M$_\odot$ \\
NB\{c,e\} & N-body & Live & Live, $\alpha,\beta,\gamma=[1.5,3,0]$ & $10^7$ & 
$2 \times 10^5$M$_\odot$ \\
NB3\{c,e\} & N-body & Live & Live, $\alpha,\beta,\gamma=[1.5,3,0^*]$ & $10^7$ 3-shell & 
$2 \times 10^5$M$_\odot^*$ \\
\hline
\end{tabular}
\end{center}
\caption[]{Simulation labels and parameters. From left to right the columns show the simulation label; a brief description (for more details see the relevant subsection in section \ref{sec:simulations}); the initial GC orbit (see section \ref{sec:gcorbit}); the background gravitational potential; the simulation resolution; and the mass of the GC ($M_c$). Parameters marked with a $*$ are allowed to vary. In section \ref{sec:gammachange}, we measure the effect of changing $\gamma$ on the NB3 simulation; in section \ref{sec:mcchange}, we measure the effect of changing $M_c$.}
\label{tab:simulations}
\end{table*}

\subsection{The semi-analytic model: SA}\label{sec:semianal}

In the semi-analytic model (SA) we solved equation \ref{eqn:starmotion} with the GC orbit held fixed (the GC initial conditions are described in section \ref{sec:gcorbit}). The underlying potential was pure harmonic and static. We used $\Omega^2 = 4/3\pi G \rho_0$ and  $\rho_0 =  9.93 \times 10^7$\,M$_\odot$\,kpc$^{-3}$. We used an isotropic, constant density, 3D, initial distribution of massless tracer particles. The equations of motion were solved using an RK4 numerical integrator \citep{1992nrca.book.....P}, with fixed timesteps of $1.5 \times 10^{-5}$\,Gyrs. This was found to conserve energy to machine accuracy over the whole simulation time in the limit $M_c \rightarrow 0$. We ran the simulations for 1\,Gyr, which is $\sim 10$ dynamical times for the GC at the core radius. This is the appropriate length of time for comparison with the full N-body run (see section \ref{sec:nbody}). We tried runs with force softening for the GC and without. There was no significant change in the results for a force softening of 10\,pc. The GC orbit was chosen to match the final stalled orbit observed in the N-body models. We ensured that the final position of the GC was identical in both models.

\subsection{The N-body model: NB}\label{sec:nbody}

In the full N-body (NB) model, we used the parallel multi-stepping N-body tree-code, pkdgrav2,
developed by \citet{2001PhDT........21S}. The potential was calculated self-consistently from the live particle distribution. The GC was allowed to freely respond to the background particles.

\begin{figure}
\begin{center}
\includegraphics[width=8cm]{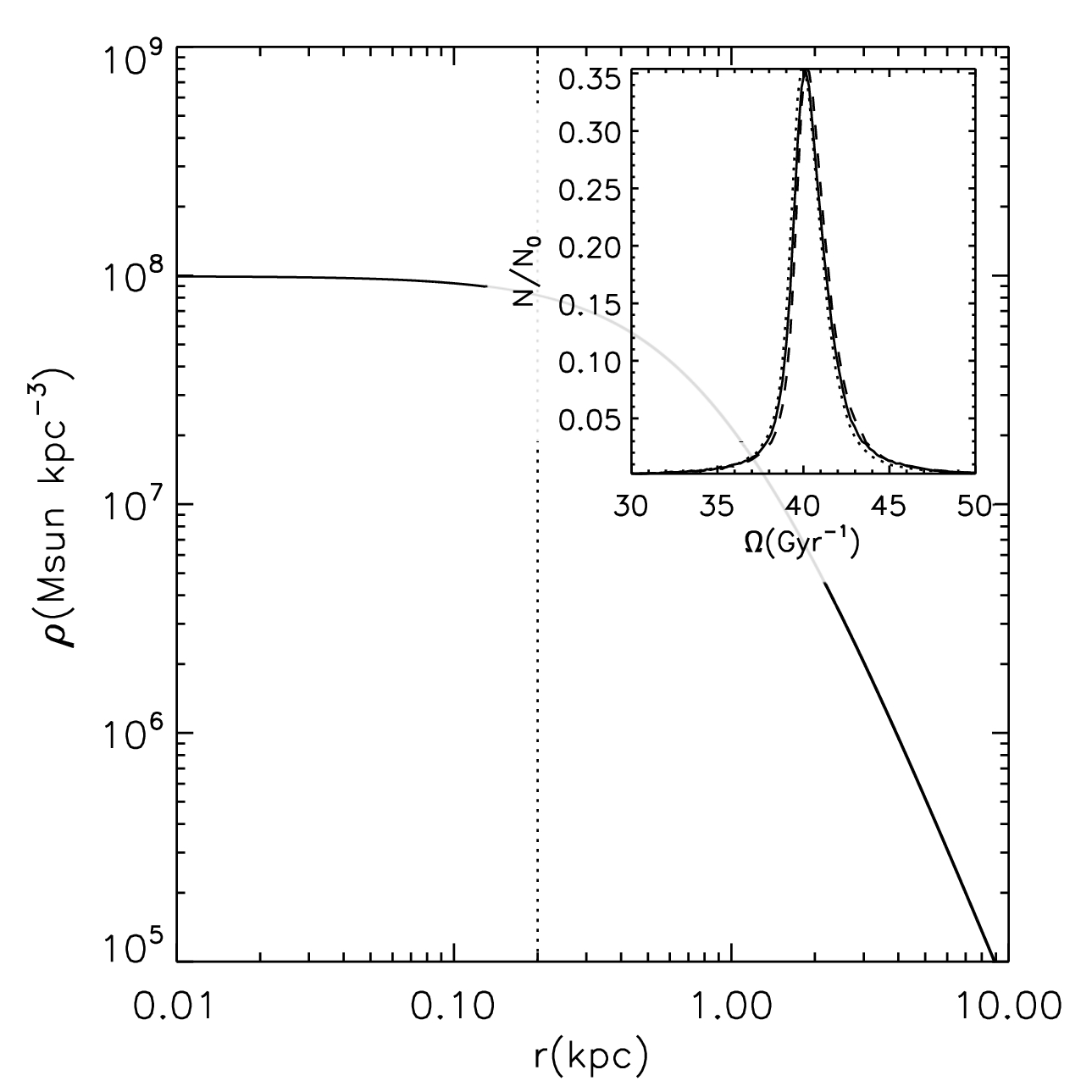}
\caption{The density distribution for the background particles used in the numerical simulations, see equation \ref{eq:cusp}. The dotted line marks the asymptotic central core where the density is constant and the potential harmonic. Inset in the plot is the distribution of orbital frequencies in the core region, plotted as $|\ddot{r}_i/r_i|$; $r_i=x_p,y_p,z_p$ (solid, dotted and dashed lines). These are equal and  strongly peaked around a single value, showing that the core is indeed harmonic.}
\label{fig:densityprofiles}
\end{center}
\end{figure}

We constructed stable particle halos using the techniques developed by \cite{2004ApJ...601...37K}. The particles are drawn self-consistently from a numerically calculated distribution function. We used a density distribution that is described by the $\alpha, \beta,
\gamma$ law (\bcite{1990ApJ...356..359H}, \bcite{1992MNRAS.254..132S}, \bcite{1993MNRAS.265..250D} and  \bcite{1996MNRAS.278..488Z}):
\begin{equation}
\rho(r)=\frac{\rho_0} {\left({\frac{r}{r_s}}\right)^\gamma \left[{1 +
\left({\frac{r}{r_s}}\right)^{\alpha}}\right]^{\frac{\beta - \gamma}
{\alpha}}}
\label{eq:cusp}
\end{equation}
where we used $\rho_0 =  9.93 \times 10^7$\,M$_\odot$\,kpc$^{-3}$, $r_s$ = 0.91\,kpc, $\alpha$ = 1.5, $\beta$ = 3.0 and $\gamma$ = 0.0. Note that $r_s$ is the {\it scale} radius, not the core radius. The radius at which the log-slope of the density profile is shallower than $-0.1$ is $r_{\mathrm{core}} \sim 200$\,pc, which defines the constant density region in this model. This halo has a virial mass of $2.0 \times 10^9$\,M$_{\odot}$ and the concentration parameter is 40. A plot of the density profile is given in Figure \ref{fig:densityprofiles}, where $r_{\mathrm{core}} $ is marked by the vertical dotted line. Inset in the plot is the distribution of orbital frequencies in the core region, plotted as $|\ddot{r}_i/r_i|$; $r_i=x_p,y_p,z_p$ (solid, dotted and dashed lines). These are equal and strongly peaked around a single value, showing that the core is indeed harmonic (c.f. equation \ref{eqn:harmpot}).

The NB run, with $10^7$ particles, corresponds to just $10^3$ particles within 300\,pc. To achieve higher resolution, in the NB3 model we also used a novel 3-shell approach \citep{marcelinprep1}. We briefly summarise this approach here, but defer the details and tests to \citet{marcelinprep1}. The 3-shell model breaks up the mass distribution into three concentric spheres. The particles in each sphere are reduced in mass and increased in number so that central regions are of higher resolution. Such a model is very useful for the current study where we would like many particles to accurately sample the central harmonic core, but are not interested in the outer density profile which may then be less accurately sampled. Massive particles from the outer sphere can and do enter the central core in this model, but they are given proportionately higher force softening to prevent them from causing spurious hard scattering. The model produces stable density profiles over $>20$\,Gyrs, very high central resolution, and no unwanted two body effects. More detailed tests are given in \citet{marcelinprep1}, but for the present study we also explicitly verified that the single component model (NB) gives comparable two-body noise (see appendix \ref{sec:twobody}).

We used a three shell model that has $10^6$ particles for the innermost sphere with 300\,pc radius, $10^6$ particles for the shell between 0.3 and 1.1\,kpc and $4 \times 10^6$ particles for the
rest of the halo. This gives us O($10^6$) particles within the core region. To achieve a similar number of particles within the central 300\,pc without the 3-shell model would require $4 \times 10^8$ particles in total. This is not yet technically feasible. Yet, as we show in section \ref{sec:noise} and appendix \ref{sec:twobody}, such high resolution is required to avoid spurious precession of the GC orbit plane. The advantage of the 3-shell model, given such limitations, is clear. The softening lengths of the particles in these shells were 3, 30 and 300 pc respectively. The particle masses were 8.9\,M$_{\odot}$, 164.0\,M$_{\odot}$ and 757.2\,M$_{\odot}$. Even the most massive particles were 100 times less massive than the GC. We experimented with varying the shell force softening and radii and found our results to be insensitive to these values.

\subsection{The GC orbit}\label{sec:gcorbit}

We used a GC mass of $M_c = 2 \times 10^5$\,M$_\odot$ with a force softening of 10\,pc. This gives $M_c/M_{en} = 0.06$, where $M_{en}$ is the total mass of background particles inside $r_\mathrm{core}$. For the N-body simulations, the GC was placed initially at a radius of 1.069\,kpc on a circular (NB3c) and elliptical (NB3e; $v_i = 0.4 v_{\mathrm{circ}}$) orbit. In both cases the GC orbited in the $x_p, y_p$ plane. For the SA simulations, the GC orbit was chosen to match that of the GC in the N-body models after it hit the constant density core and stalled. Its orbital phase was chosen such that at the end of the SA simulation (after 1\,Gyr) the GC would be in the same place as in the N-body simulations.

\subsection{Particle noise, resolution and convergence}\label{sec:noise}

We had surprising difficulty in obtaining enough resolution in the N-body simulations for our results to be believable. The problem centred around the precession of the GC orbit plane. For a spherical potential (such as that studied here) all orbits, including that of the GC, should be planar. However, in our initial lower resolution runs, with a resolution of $10^5$ particles within 300\,pc, we found that the GC orbit plane would precess, sometimes by as much as 20$^{\mathrm{o}}$ over 10\,Gyrs. Since we are trying to model an effect that relies critically on the orientation of the GC orbit plane, it is essential that the plane remains stable.

In appendix \ref{sec:twobody}, we use a simple analytic model of a 2D random walk to prove that this precession is a result of two-body noise in the simulations. Reducing such noise drove us to use the three shell model discussed above. We show in appendix \ref{sec:twobody}, that the noise is not some special property of the three shell model, but is present in all N-body simulations. We found that some initial GC orientations showed more precession than others, for the same resolution. This is to be expected from a random walk driven by two-body noise. The effect of such precession was found to be quite small. However, it does lead to a spurious (and very slow, sub-Chandrasekhar) decay of the GC orbit once it reaches the core. We present the results here from simulations which showed the minimal GC plane precession. However, our main results are not sensitive to such selection. Nor are our results sensitive to the use of the three shell model.

\section{Results}\label{sec:results}

\subsection{The stalling of dynamical friction in the core}\label{sec:stall}

\begin{figure}
\begin{center}
\includegraphics[width=8cm]{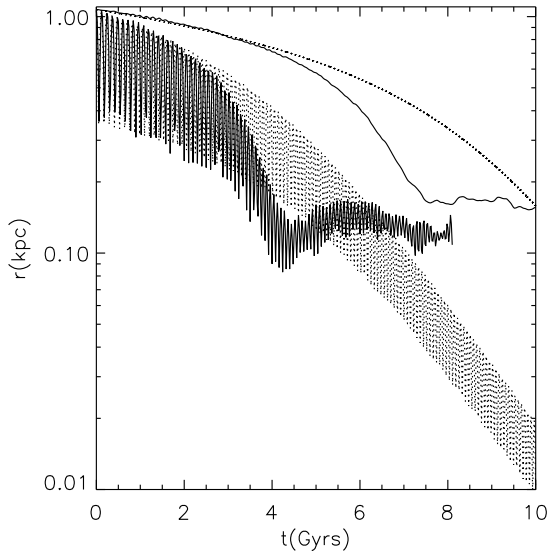}
\caption{The decay of the radius of the GC as a function of time for a GC on a circular (straight solid line) and elliptical (oscillating solid line) orbit. Overlaid (dotted lines) are the predictions from the Chandrasekhar formula given in equation \ref{eqn:chandrasekhar}, using $\ln\Lambda=5$. Notice that for the first few Gyrs the agreement with equation \ref{eqn:chandrasekhar} is excellent. As the cluster nears the constant density core ($r_{\mathrm{core}}\sim 200$\,pc), it enters a phase of super-Chandrasekhar dynamical friction, after which dynamical friction practically ceases.}
\label{fig:decay}
\end{center}
\end{figure}

Figure \ref{fig:decay} (straight solid line) shows the decay of the radius of the GC as a function of time for the NB3c and NB3e simulations (see Table \ref{tab:simulations}). Overlaid is the prediction from the Chandrasekhar formula given in equation \ref{eqn:chandrasekhar}. For this we used a {\it constant} $\ln\Lambda = 5$, which is the value we use throughout this paper. If we equate $b_\mathrm{min}$ with the GC force softening, $b_\mathrm{min} = 10$\,pc, this gives $b_\mathrm{max} \simeq 1.5$\,kpc, which is of order the `size' of our system. This is consistent with values found in other numerical studies of dynamical friction on point mass particles (see e.g. \bcite{2003MNRAS.344...22S}). 

As the cluster nears the constant density core ($r_{\mathrm{core}} \sim 200$\,pc), it enters a phase of super-Chandrasekhar dynamical friction, after which dynamical friction practically ceases. This occurs irrespective of the initial GC orbit. This is in excellent qualitative agreement with analytic expectations from section \ref{sec:analytic}. 

\begin{figure*}
\begin{center}
\includegraphics[width=6.5cm]{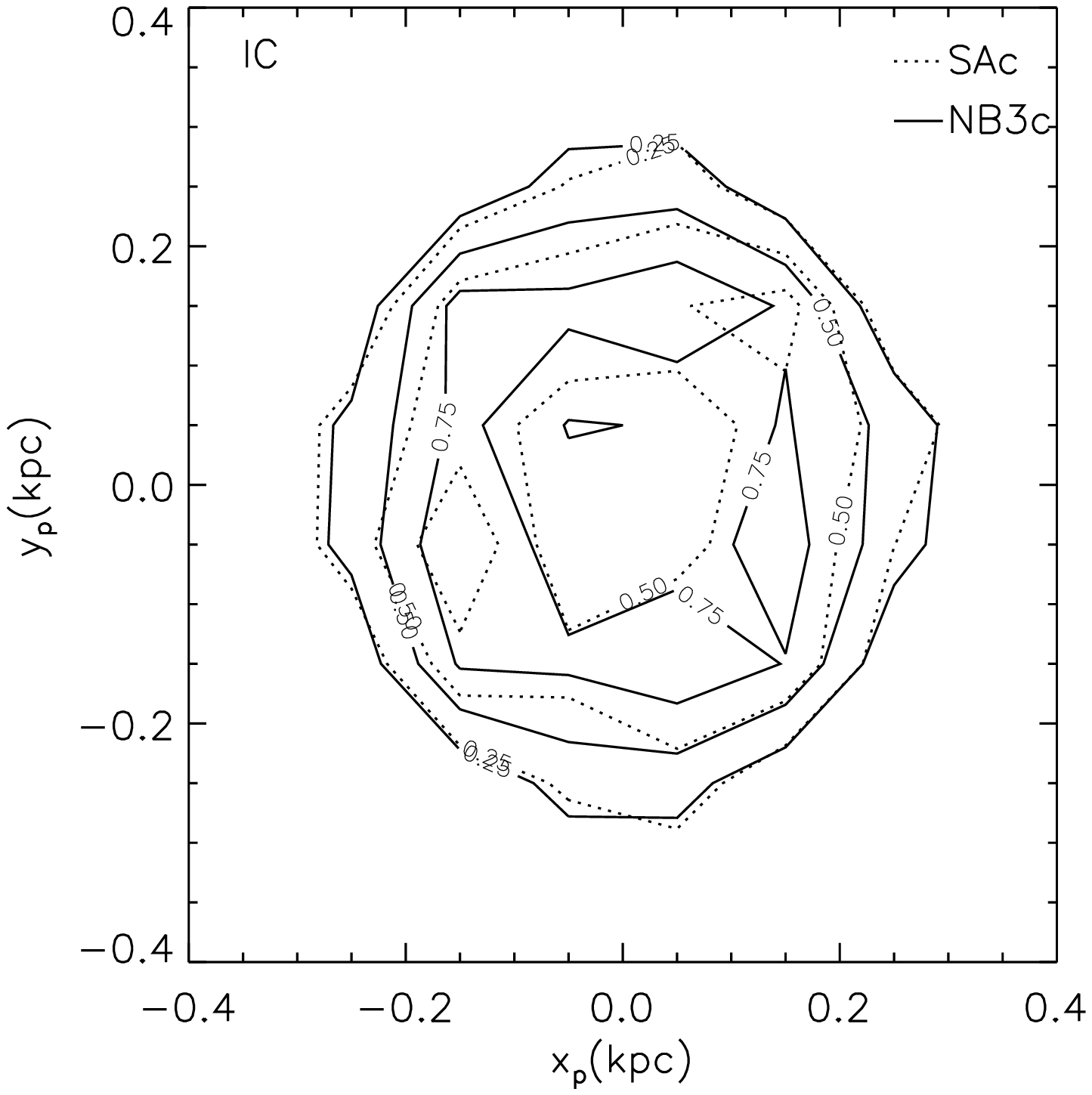}
\includegraphics[width=6.5cm]{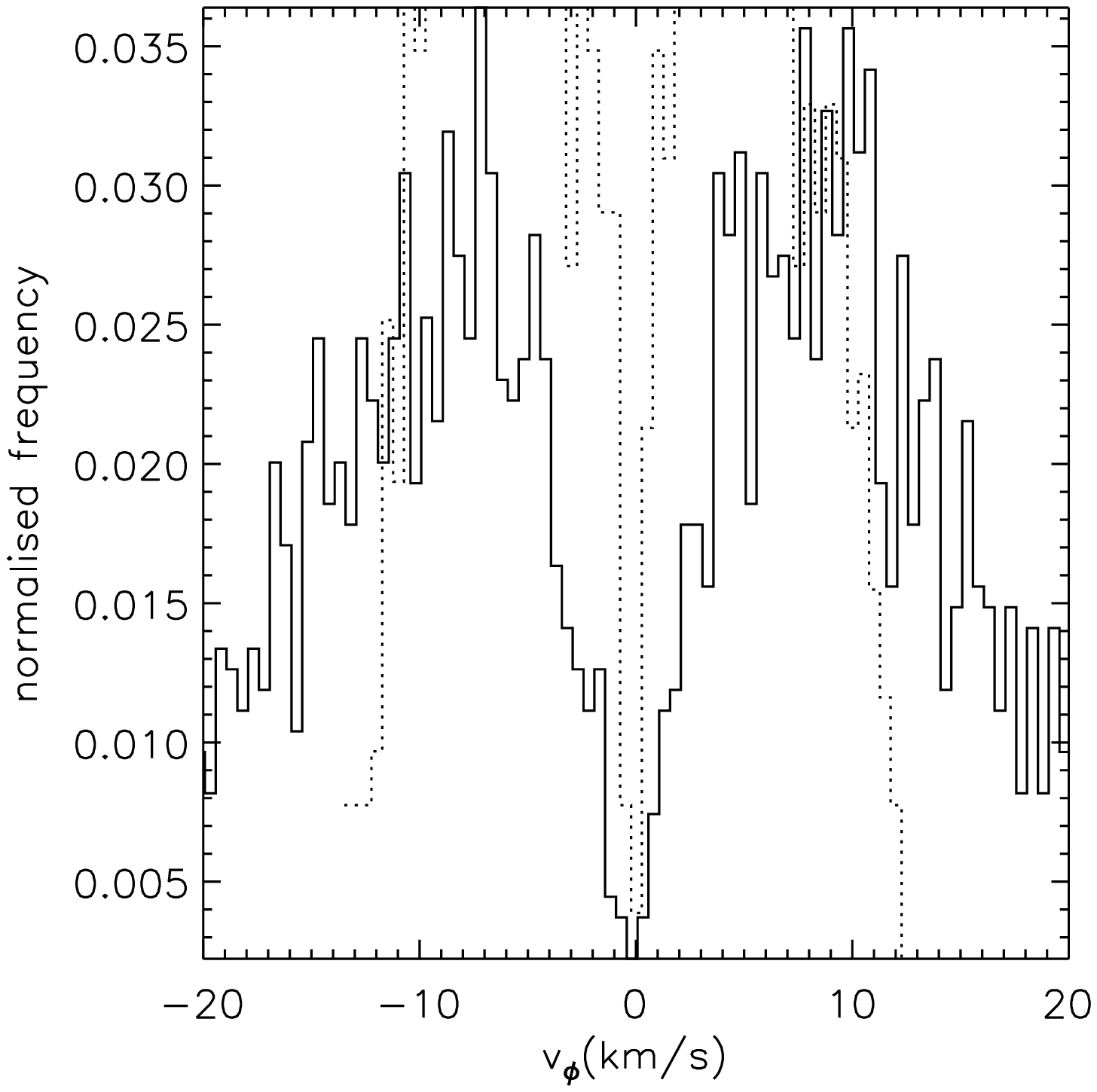}\\
\includegraphics[width=6.5cm]{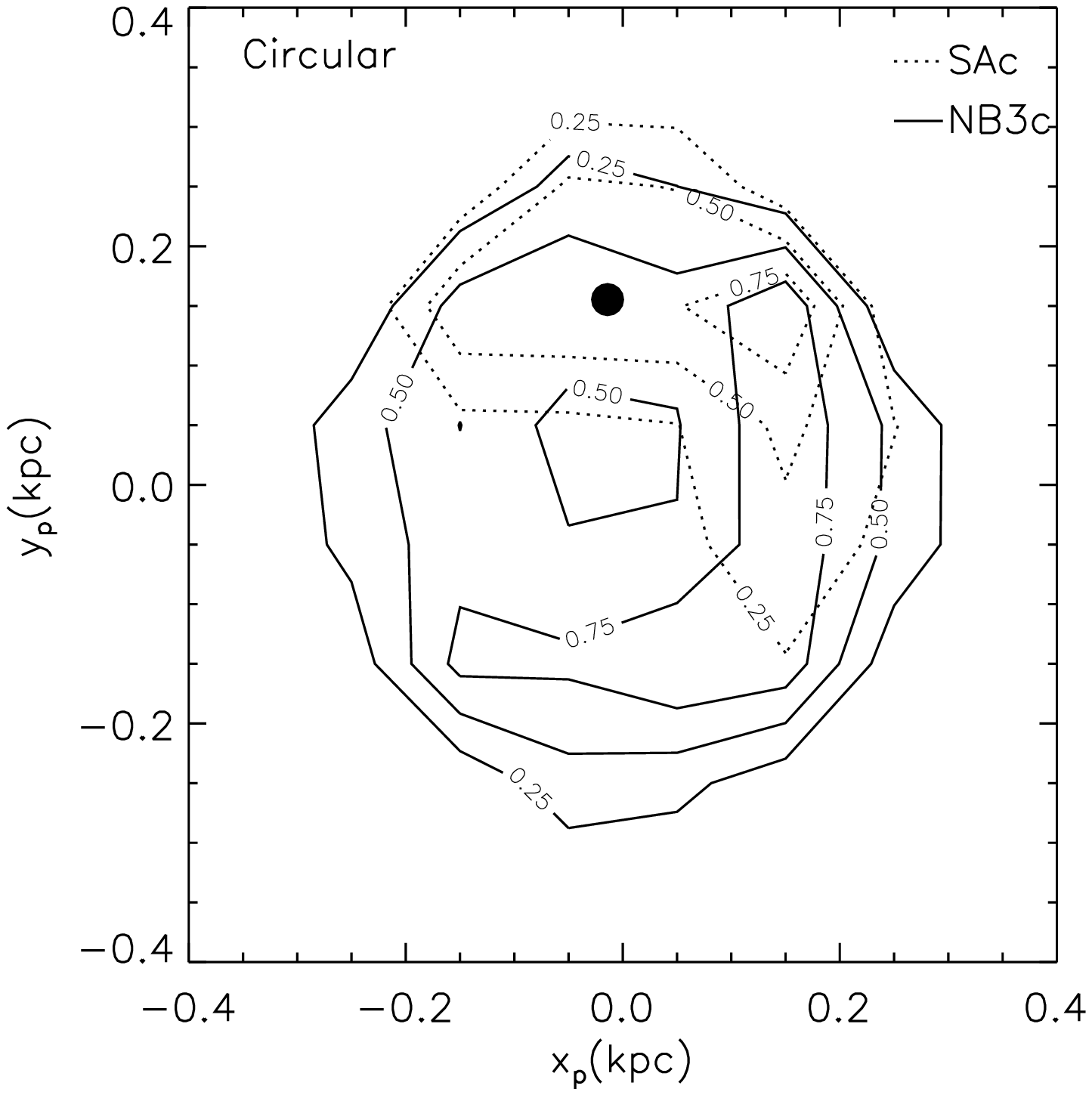}
\includegraphics[width=6.5cm]{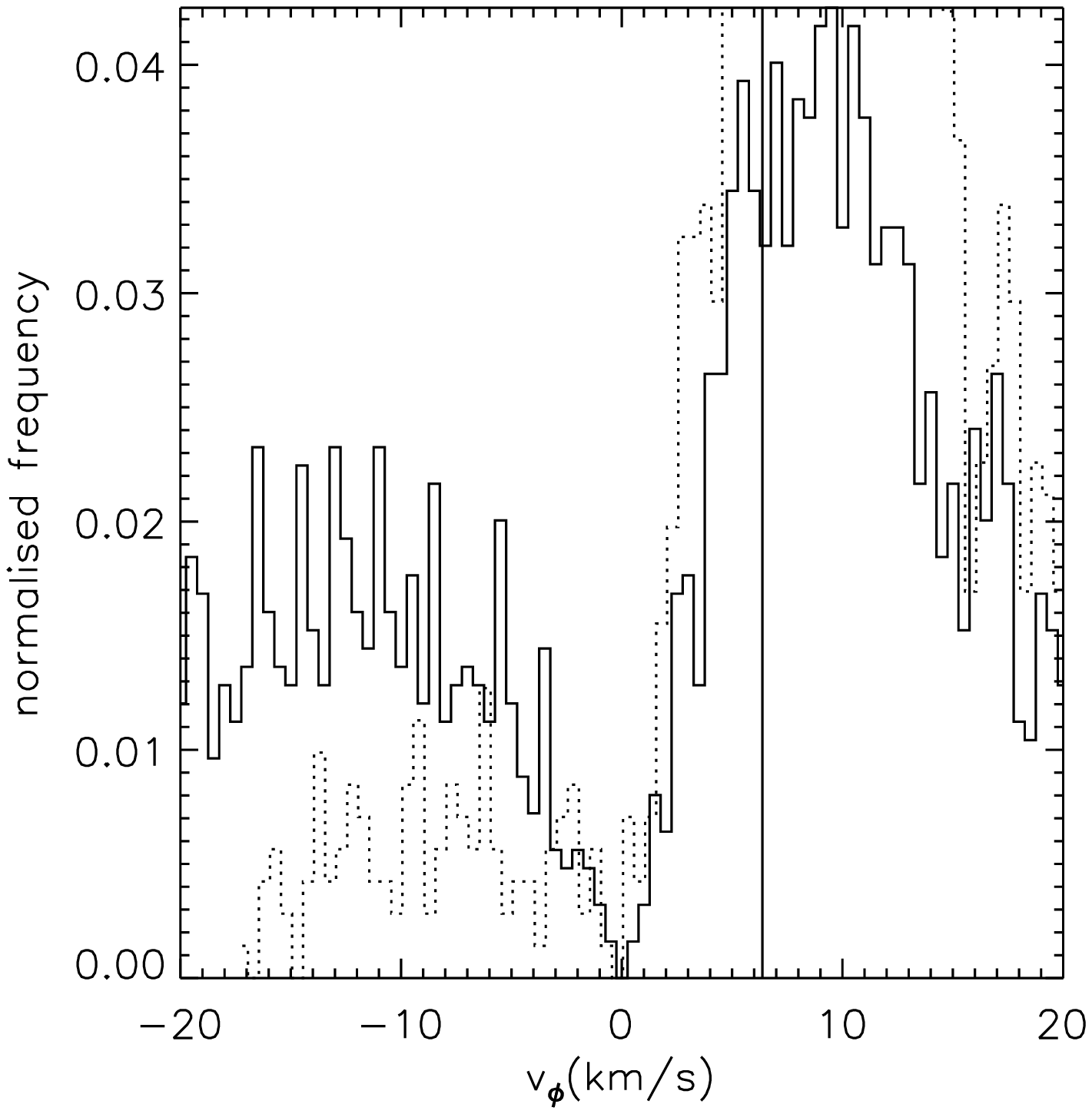}\\
\includegraphics[width=6.5cm]{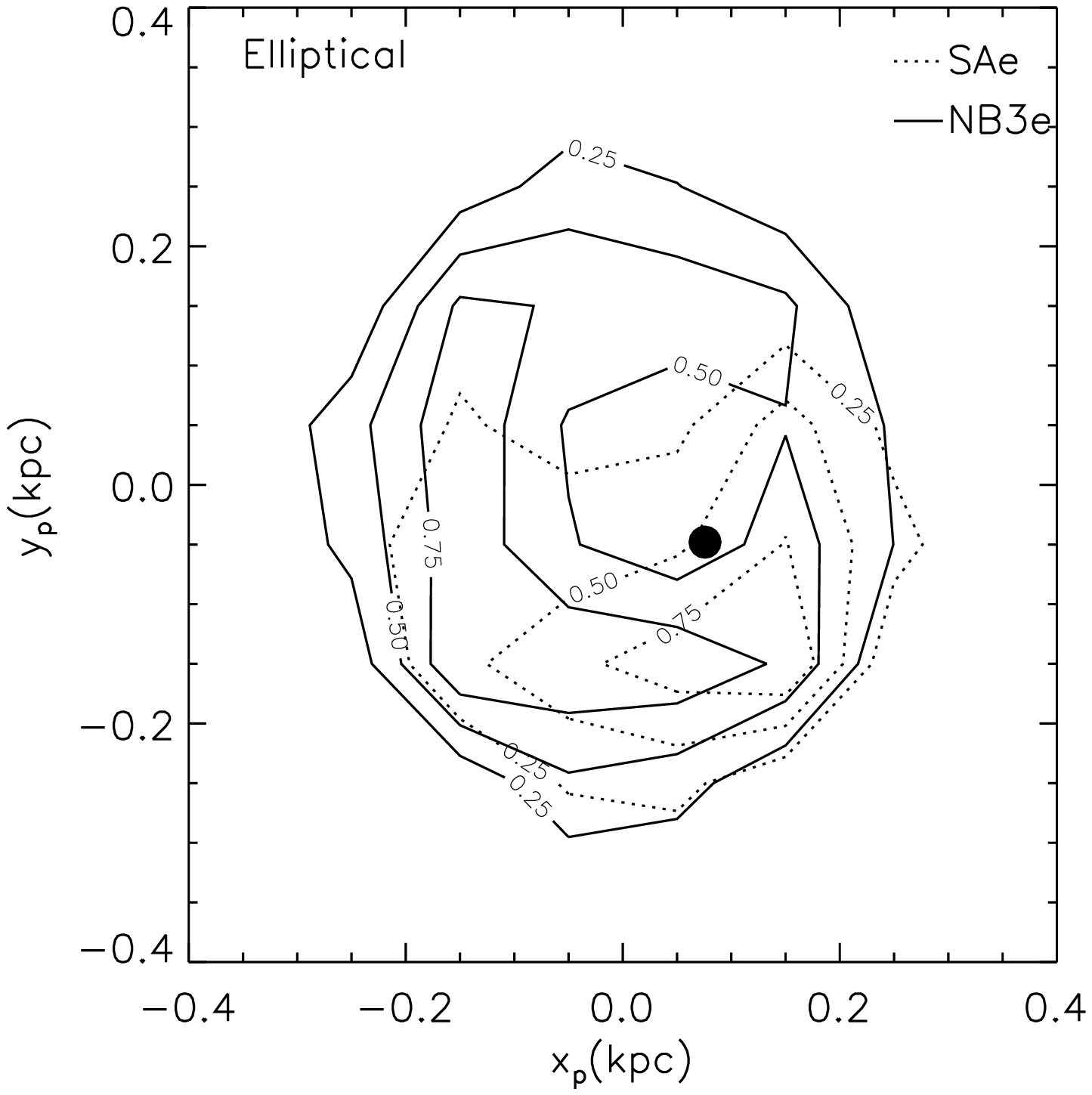}
\includegraphics[width=6.5cm]{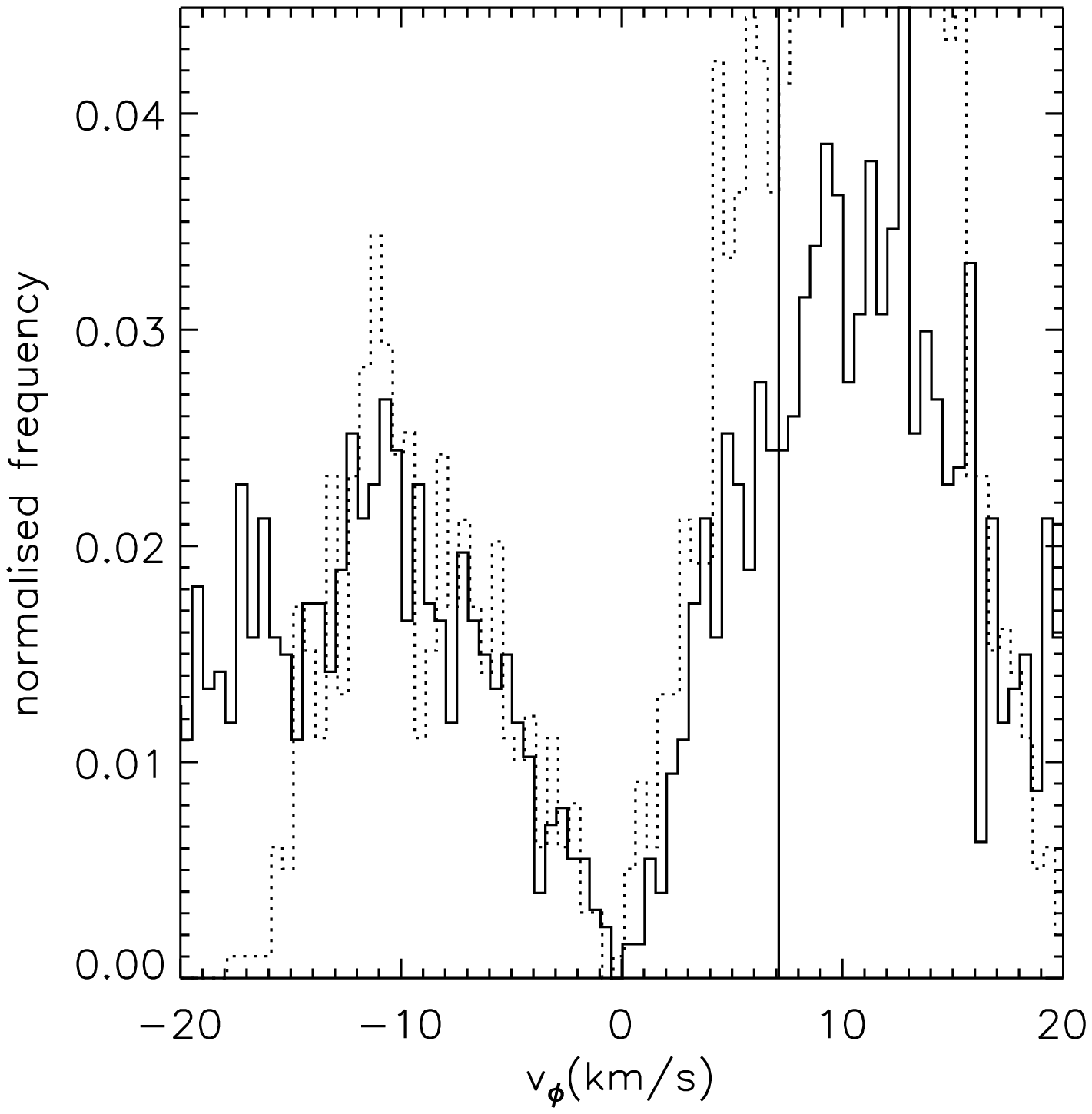}
\caption{The distribution of particles in the $x_p,y_p$ plane for the N-body (solid lines) and semi-analytic (dotted lines) simulations. The left panels show density contours for the particle distribution (that was initially constant-density) in the $x_p,y_p$ plane. The right panels show velocity histograms for the $v_\phi$ component of the velocity; where $v_\phi$ is the velocity about the $z_p$-axis. The solid vertical line marks the GC velocity about the $z_p$-axis. The top two panels show the GC circular orbit simulations just before the GC experiences super-Chandrasekhar dynamical friction. This corresponds to $t=0$\,Gyrs for the SAc simulation, and $t=5$\,Gyrs for NB3c. Notice that, for the NB3c simulation (solid lines), the background particles are nearly unchanged from their initial distribution. The middle two panels show the GC on a circular orbit after the super-Chandrasekhar friction has finished and the GC has settled into a steady-state in the harmonic core. This corresponds to $t=1$\,Gyrs for the SAc simulation, and $t=8$\,Gyrs for NB3c. The bottom two panels show the GC on an elliptical orbit ($v_i = 0.4 v_{\mathrm{circ}}$) after it has reached the harmonic core. This corresponds to $t=1$\,Gyrs for the SAe simulation, and $t=4$\,Gyrs for NB3e. In all cases the final position of the GC in the SAc,e and NB3c,e simulations is identical and marked by the solid circle. Note that there is no GC marked in the top panels since, in the NB3c simulation, the GC lies outside of the plot area at this time.}
\label{fig:results}
\end{center}
\end{figure*}

Figure \ref{fig:results} shows the distribution of particles in a slice about the orbit plane of the GC. The slice is defined such that  $|\underline{J}_p\cdot\underline{J}_c| < |\underline{J}_p| |\underline{J}_c|\cos(\theta)$, with $\theta = 10^{\mathrm{o}}$, where $\underline{J}_{p,c}$ is the specific angular momentum of the particle and GC respectively. The left panel shows density contours for the particle distribution (which was initially constant-density) in the $x_p,y_p$ plane. The right panel shows velocity histograms for the $v_\phi$ component of the velocity; where $v_\phi$ is the velocity about the $z_p$-axis. We do not show the $v_r$ and $v_\theta$ components of the velocity, since they are not altered from the initial conditions and remain approximately Gaussian ($r,\theta$ and $\phi$ are the usual spherical polar coordinates). In the top panels, the solid lines show the slice just before the GC hits the core in the NB3c simulation (at time $t=5$\,Gyrs); the dotted contours show the SAc simulation at $t=0$. The middle panels show similar results for the NB3c and SAc simulations at times $t=8$\,Gyrs and $t=1$\,Gyrs, respsectively. The bottom panels show the NB3e and SAe simulations at times $t=4$\,Gyrs and $t=1$\,Gyrs, respectively. We analyse within this slice to highlight the changes in density caused by the GC. Outside of the slice, background particles still move on epicycles about the GC, but their projected positions onto the $x_p,y_p$ plane make it difficult to see the density enhancement about the GC.

Notice that the velocity histograms for $v_\phi$ (right panel, top) are double-peaked. This is because we have taken a thin slice in the orbit-plane of the cluster. The only particles that have $v_\phi =0$ in this plane are those on pure radial orbits, which is a very small number of particles in the isotropic initial conditions.

In the top panels of Figure \ref{fig:results}, the particles are close to their initial configuration. There has been some depletion in density at the centre and the on-set of some substructure, but the velocity histograms show that the velocity distribution of the background particles is still isotropic. 

The middle and bottom panels in Figure \ref{fig:results} show the distribution of particles in the slice after the super-Chandrasekhar friction has ended, and the GC has settled into equilibrium in the core. Notice the good agreement with the semi-analytic (SAc,e) simulations for both the density and velocity distribution in the slice, irrespective of the initial GC orbit. As expected from the arguments given in section \ref{sec:analytic}, the number of counter-rotating particles has been significantly depleted. 

The density distribution in the slice is peaked just behind the cluster; it has a tail which is longer for the full N-body run. This is likely due to particle-particle scattering which prevents high density regions from forming. Such a tail should lead to some dynamical friction on the GC from the background. We estimated the strength of this effect for the SAc model. To do this, we summed the force from all of the background particles on the GC, assuming that their total mass was $M_{en}$. What really matters is the {\it time averaged} force on the GC. This must be small, since little or no dynamical friction is observed after the GC reaches the core. However, even the total force at an instant is always smaller than the dynamical friction force, computed from equation \ref{eqn:chandrasekhar}.

Notice from equation \ref{eqn:r}, that we could construct {\it any} final density distribution using an appropriate combination of epicyclic orbits about the GC. In practise, however, the final density distribution is set by the initial configuration of background particles within the core. The transformation of this initial distribution by the arrival of the GC, must be determined numerically. The SA model is essential in this respect.

The keen observer will notice that the enhanced friction appears to set in rather near the region where the resolution in the 3-shell model increases (recall that the high resolution inner shell starts at 300\,pc). This is almost certainly a coincidence. We performed two tests to check this. Firstly, an explicit test by starting a GC sinking {\it inside} the high resolution shell. Once again, we observed enhanced friction followed by stalling. Secondly, we performed a test-run starting the GC outside the 2nd shell. As it moved through the shell transition at 1.1\,kpc, no detectable effect was observed.

\subsection{The effect of varying $M_c$}\label{sec:mcchange}

\begin{figure}
\begin{center}
\includegraphics[width=8cm]{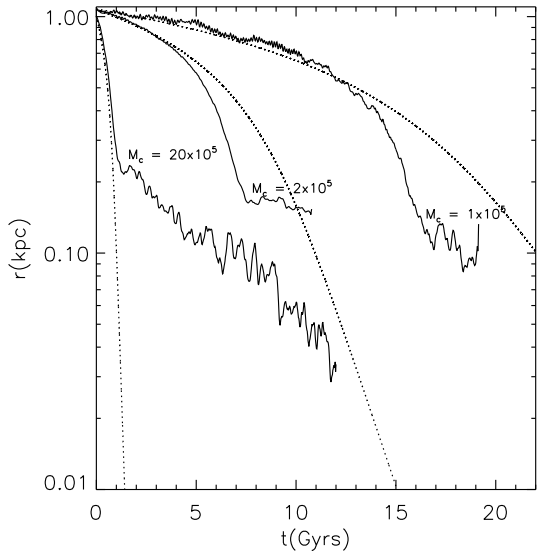}
\caption{The decay rate of the GC as a function of $M_c$ (solid lines); $M_c$ is marked in solar masses. Overlaid are analytic predictions from equation \ref{eqn:chandrasekhar} using $\ln\Lambda = 5$, as previously (dotted lines). For these simulations, we re-ran simulation NB3c (see Table \ref{tab:simulations}) but using 10 times the GC mass ($M_c = 2\times 10^6$M$_\odot$), and half of the GC mass ($M_c = 1\times 10^5$M$_\odot$). $M_{en}/M_{c}$ at the point of the onset of the stalling behaviour is, in order of increasing $M_c$: [3,5,2].}
\label{fig:varyingmc}
\end{center}
\end{figure}

Figure \ref{fig:varyingmc} shows the decay rate of the GC as a function of the GC mass, $M_c$ (solid lines); $M_c$ is marked in solar masses. Overlaid are analytic predictions from equation \ref{eqn:chandrasekhar} using $\ln\Lambda = 5$, as previously (dotted lines). For these simulations, we re-ran simulation NB3c but using 10 times the GC mass ($M_c = 2\times 10^6$M$_\odot$), and half of the GC mass ($M_c = 1\times 10^5$M$_\odot$). For reasons of computational expense, we ran these new runs at a lower resolution with $10^5$ particles in each shell. We could not investigate smaller GC masses than $M_c = 10^5$M$_\odot$, since then $M_c$ approaches the mass of the heaviest particle and two-body effects dominate over dynamical friction.  

Notice that the lower resolution runs are noisier and decay faster once the GC hits the core. This decay is due to the precession (due to numerical noise) of the GC orbit plane, discussed in section \ref{sec:noise} and appendix \ref{sec:twobody}, and is much smaller in the higher resolution runs. We explicitly checked that this is indeed the case using lower resolution runs of NB3c. 

In all runs, the GC shows a reduced friction at the core region. Notice that the point at which the GC departs from Chandrasekhar like friction appears to be a weak function of the GC mass. This is to be expected: a more massive GC will more rapidly scatter the background particles and stall more quickly once it reaches the core region. However, it is tempting to suggest a simpler explanation: that equation \ref{eqn:chandrasekhar} is failing simply because $M_{en} = \eta M_c$; where $M_{en}$ is the final mass enclosed and $\eta$ is some constant of order unity. This is perhaps worrisome given the mass ratios, $M_{en}/M_{c}$, at the point of the onset of the stalling behaviour. These are, in order of increasing $M_c$: [3,5,2]. However, we believe that the situation is not this simple for the following reasons: (i) if the stalling were a result only of $M_c \simeq M_{en}$, then it would not be a special property of constant density cores. We show in section \ref{sec:gammachange}, below, that the stalling behaviour does not occur for steeper density profiles, whatever the enclosed mass. (ii) The model we present in section \ref{sec:analytic} provides a good fit to the final density and velocity distribution of the background particles in the core suggesting that we have captured the correct physical explanation. 

\subsection{The effect of varying $\gamma$}\label{sec:gammachange}

\begin{figure}
\begin{center}
\includegraphics[width=8cm]{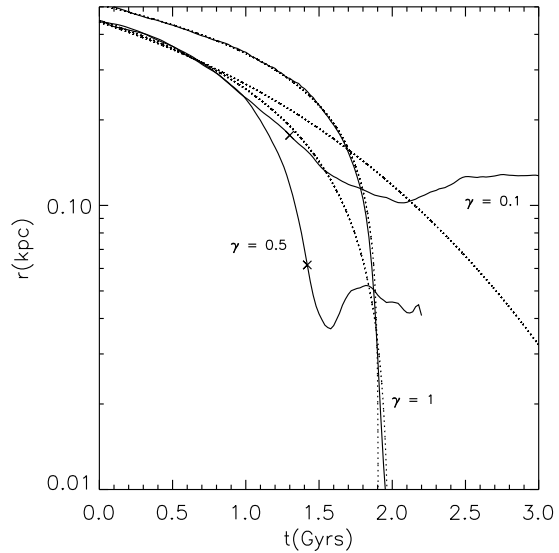}
\caption{The decay rate of the GC as a function of the central log-slope of the background density distribution, $\gamma$ (solid lines); $\gamma$ is marked on the plot. Overlaid are analytic predictions from equation \ref{eqn:chandrasekhar} using $\ln\Lambda = [8,7,3.5]$, in order of increasing $\gamma$ (dotted lines). The crosses mark the radii at which the {\it final} density profile has a central log-slope shallower than -0.1. $M_{en}/M_{c}$ at the point of the onset of the stalling behaviour is, in order of increasing $\gamma$: [4,1].}
\label{fig:varyinggamma}
\end{center}
\end{figure}

Figure \ref{fig:varyinggamma} shows the decay rate of the GC as a function of the central log-slope of the background density distribution, $\gamma$ (solid lines); $\gamma$ is marked on the plot. For these simulations, we re-ran simulation NB3c but using $\gamma = [0.1,0.5]$. Also shown are results for a simulation with $\gamma=1$ taken from \citet{Goerdt:2006rw}. Overlaid are analytic predictions from equation \ref{eqn:chandrasekhar} using $\ln\Lambda = [8,7,3.5]$, in order of increasing $\gamma$ (dotted lines). $\ln\Lambda$ is different for each of these simulations, reflecting the change in the underlying density distribution; similar results have been found elsewhere in the literature (see e.g. \bcite{2005A&A...431..861J}). All simulations were high resolution ($\sim 10^6$ particles per shell), but since we are interested in the core stalling properties of the GC, we started the $\gamma = [0.1,0.5]$ simulations at $\sim 400$\,pc, rather than $\sim 1$\,kpc as previously. 

The key point is that the $\gamma=1$ model is well-fit by the Chandrasekhar form over the entire simulation time. This is despite the fact that $M_{en} \simeq M_c$ at $\sim 0.1$\,kpc for this run. This suggests that the core stalling behaviour is a special property of the harmonic core and not to do with the enclosed mass. However, the $\gamma = [0.1,0.5]$ runs both show stalling behaviour despite not having a central core. This occurs because the GC itself {\it creates} a small core as it falls in and heats the background particle distribution. For initial density distributions steeper than $\gamma=0.5$ this no longer occurs. In this case, the density profile does become shallower as a result of heating, but the heating is not sufficient to form a core in the centre before the GC falls all of the way in. The crosses on Figure \ref{fig:varyinggamma} mark the radii at which the final density profile has a central log-slope shallower thant -0.1. Recall that this is the same definition we used to define $r_\mathrm{core}$ earlier. 

\section{Conclusions}\label{sec:conclusions}

Using analytic calculations and N-body simulations we have shown that in
constant density harmonic cores, sinking satellites undergo an initial
phase of very rapid (super-Chandrasekhar) dynamical friction, after
which they experience no dynamical friction at all. This occurs because, for the special case of harmonic potentials, there are stable solutions where the background particles move on epicycles about the in-falling satellite. The system moves rapidly into this stable configuration. In doing so, the satellite experiences a brief moment of enhanced friction. Once in equilibrium, there is no net momentum transfer between the background particles and the satellite and friction ceases. For density profiles with a central power law profile, $\rho \propto r^{-\alpha}$, the infalling satellite heats the background and causes $\alpha$ to decrease. For $\alpha < 0.5$ initially, the satellite generates a small central constant density core and stalls as in the $\alpha = 0$ case. 

Our results concerning dynamical friction stalling in constant density cores are of broad astrophysical interest. Recent observational work suggests that galaxies may have central dark matter density cores, rather than the $r^{-1}$ density cusps predicted by numerical simulations. Galactic bars orbiting in such potentials will experience very weak dynamical friction and can be very long-lived (in fact central density distributions do not need to be pure harmonic to see this effect, low-density will also lead to very little friction -- see e.g. \bcite{1998ApJ...493L...5D} and \bcite{2000ApJ...543..704D}). Satellites falling into such galaxies will stall at the core radius and never make it to the centre. This point was investigated in a companion paper \citep{Goerdt:2006rw}, where we suggested that a constant density core could solve the `timing problem' for the GCs in the Fornax dwarf galaxy. Finally, recent work on merging black holes suggests that they can form a central constant density core in the background distribution (see e.g. \bcite{2002MNRAS.331L..51M} and \bcite{2002ApJ...566..801R}) prior to forming a hard binary. If true, our results suggest that this could further exacerbate the well-known problem of getting the binaries to coalesce. Their rate of hardening will stall, even before the majority of stars and dark matter have been ejected from the core, if the background distribution is close to constant density. This may point towards gas playing a more important role in bringing supermassive black holes together at the centre of galaxies (see e.g. \bcite{2000ApJ...532L..29G}).

\section{Acknowledgements}
We would like to thank Lucio Mayer, Victor Debattista, Scott Tremaine, Prasenjit Saha, Andrey Kravtsov,  and the anonymous referee for useful comments which greatly aided the clarity of this work. AP is supported by a PPARC studentship. Special thanks go to Doug Potter, without whom none of this would have been possible. He has been the wizard of the zBox2 on which all of the simulations presented here were performed. 

\appendix
\section{Two body noise and precession of the GC orbit plane}\label{sec:twobody}

In this appendix we present a simple analytic model for the precession of the GC orbit plane due to particle noise and compare this with the simulations. We show that even quite small particle noise can lead to significant plane precession over $\sim 100$ dynamical times. 

Under the assumption of linear background particle trajectories, it is straightforward to show that an interaction with one background particle will produce a velocity kick perpendicular to the GC's orbit plane given by \citep{1987gady.book.....B}:

\begin{equation}
\delta v_z = \frac{2mbv^3}{G(M_c+m)^2}\left[1+\frac{b^2v^4}{G^2(M_c+m)^2}\right]^{-1}
\label{eqn:vkick}
\end{equation}
where $m$ is the mass of the background particle, $M_c$ is the mass of the GC and $b$ is the impact parameter (initial perpendicular separation) of the encounter. Such a kick occurs over $\sim$ a dynamical time. 

\begin{figure}
\begin{center}
\includegraphics[width=8cm]{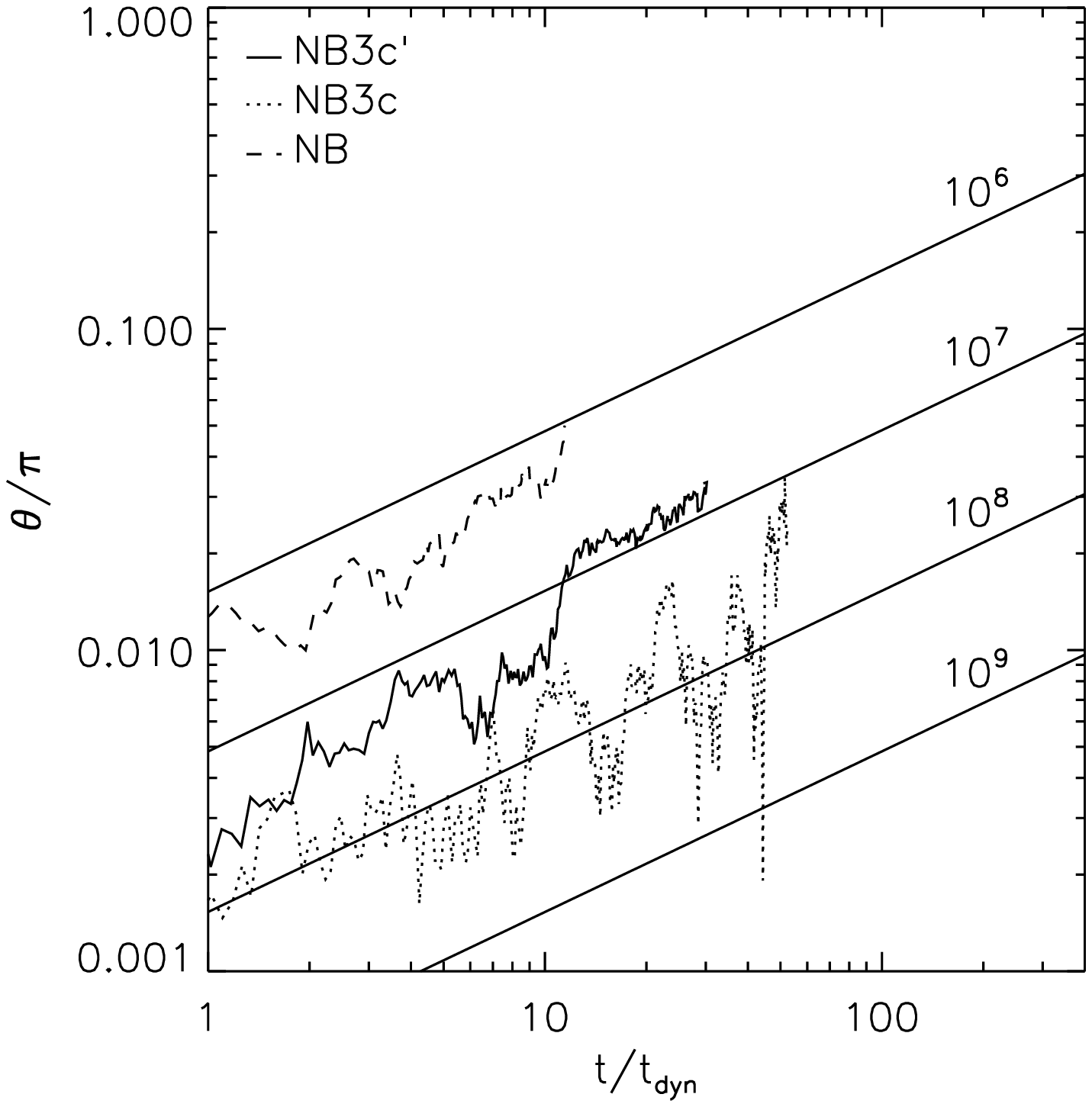}
\caption{Evolution of the GC orbit plane in angle, $\theta$, over the simulation time. The straight solid lines are for an analytic model that assumes a 2D random walk. Results are shown for increasing particle number, $N$. Over-plotted are results from three typical N-body simulations: NB, NB3c and NB3c' - see Table \ref{tab:simulations} and this appendix for details; NB3c' is identical to NB3c, except that the GC initial orbit plane is different.}
\label{fig:randomwalk}
\end{center}
\end{figure}

Summing over all such encounters (all impact parameters) then gives the mean total velocity kick to the GC in $\sim$ a dynamical time. We sum over $\delta v_z^2$ to give the r.m.s. change; $\delta v_z$ is of random sign and will sum to zero:

\begin{eqnarray}
\frac{\Delta v_z^2}{v^2} & = & \frac{1}{v^2}\int_{b_\mathrm{min}}^{b_\mathrm{max}} \delta v_z^2  \frac{2Nb}{b_\mathrm{max}^2}db\nonumber \\
& = & \frac{8}{2N}\left[\frac{x_\mathrm{min}^2-x_\mathrm{max}^2}{(1+x_\mathrm{min}^2)(1+ x_\mathrm{max}^2)}+\ln\left(\frac{1+x_\mathrm{max}^2}{1+x_\mathrm{min}^2}\right)\right] \nonumber \\
& = & \frac{8}{N}\ln\Lambda'
\label{eqn:perpkick}
\end{eqnarray}
\begin{equation}
x_\mathrm{max} = \frac{Nm}{M_c+m}\hspace{1mm} ; \hspace{1mm}x_\mathrm{min} =  \frac{Nm}{M_c+m}\Lambda^{-1}
\end{equation}
where $v$ is now $\sim$ the velocity of the GC; $\Lambda = b_\mathrm{max}/b_\mathrm{min}$ is the term inside Coulomb logarithm that also appears in equation \ref{eqn:chandrasekhar}; $N$ is the number of background particles inside $\sim b_\mathrm{max}$ (the GC orbit is assumed to lie in the x-y plane); and $\ln\Lambda' = 1/2[...]$ is defined by equation \ref{eqn:perpkick}. 

Notice that in the limit of large impact parameters, $b_\mathrm{max} \gg b_\mathrm{min} \gg GM_c/v^2 \Rightarrow x_\mathrm{max} \gg x_\mathrm{min} \gg 1$, and equation \ref{eqn:perpkick} reduces to the more familiar form: $\Delta v_z^2/v^2 = (8/N) \ln\Lambda$. It is then independent of the GC mass. 

In one orbit, the GC will move a mean $z$ distance, $\Delta z \sim \Delta v_z t_\mathrm{dyn}$, where $t_\mathrm{dyn}$ is the orbit time. The mean change in angle over one orbit, $\Delta \theta$, of the vector normal to the GC orbit plane is then given by:

\begin{equation}
\Delta \theta \sim \tan^{-1}\left(\frac{\Delta z}{r}\right) = \tan^{-1}\left(2\pi \sqrt{\frac{8\ln \Lambda'}{N}}\right) 
\label{eqn:deltatheta}
\end{equation}
\noindent
where we have assumed that the GC moves on a circular orbit of radius, $r$.

Any dependence on the underlying potential completely factors out in equation \ref{eqn:deltatheta}, and $\Delta \theta$ depends only on the number of particles, $N$, and very weakly on $M_c/m$.

The orbit plane can be tilted due to such scattering noise from the background distribution in two independent directions. Since the potential is spherical, there is no restoring force and once the plane has tilted, the probability it will tilt again is independent of its past history. Thus we may model the accumulated precession of the orbit plane by a 2D random walk. This gives:

\begin{equation}
\theta = \Delta \theta \sqrt{\frac{t}{t_\mathrm{dyn}}}
\label{eqn:randomwalk}
\end{equation}

The orbit time at $r = r_\mathrm{core}$ for our model is given by $t_\mathrm{dyn} = 2\pi \sqrt{\frac{r^3}{GM(r)}} = 0.15$\,Gyrs. In Figure \ref{fig:randomwalk} we plot the mean orbit plane precession predicted by this random walk model, as a function of simulation time, $t/t_\mathrm{dyn}$. We use $b_\mathrm{max} = 1.5$\,kpc and $b_\mathrm{min} = 10$\,pc, which gives $\ln\Lambda = 5$, as in section \ref{sec:results}. In section \ref{sec:results}, we typically ran our N-body models for 10\,Gyrs which corresponds to $\sim 100$ dynamical times. The straight solid lines show the effect of increasing the particle number, $N$. Notice that extremely high resolution is required to keep plane precession to a minimum over our simulation time: even with $10^7$ particles we can expect a mean precession over the whole simulation of $\sim 7^{\mathrm{o}}$. Over-plotted are results from the NB, NB3c and NB3c' simulations (see Table \ref{tab:simulations}). Recall that the NB model was a single shell model with $10^7$ particles in total, with $10^3$ within 300\,pc. The NB3 simulations were three-shell models with $10^6$ particles within 300\,pc. NB3c' is a simulation which is identical to NB3c but with a different GC initial orbit plane. Notice that in all cases the plane precesses; it is not some numerical error introduced by the three shell model. The NB3 simulations show a smaller precession than the NB simulation as is expected given their higher effective resolution. Finally, notice that changing the initial GC orbit plane can alter the total precession quite dramatically (compare the NB3c and NB3c' simulations). This is to be expected given the random walk model, above.

The total particle number, $N$, in equation \ref{eqn:deltatheta} is a slightly ill-defined quantity and so should not be equated exactly with the number of particles in the simulation (particularly for the 3-shell models). However, it is encouraging that our simple random walk model produces the correct mean slope for the plane precession and the correct scaling with particle number. It is clear that the 3-shell model has an advantage over the single shell model: it samples the core region with 1000 times the resolution of the single shell and shows much smaller two-body noise.

Throughout this paper, we present simulations which minimise the evolution of the orbit plane. It is important to note, however, that {\it all} of our simulations show the same central result: a period of super-Chandrasekhar friction, followed by stalling at the constant density core. 

\bibliographystyle{mn2e}
\bibliography{/Users/justinread/Documents/LaTeX/BibTeX/refs}
 
\end{document}